\documentclass[twocolumn,trackchanges]{aastex701}
\usepackage{amsmath}
\usepackage{multirow}

\begin{document}

\title{Time-Domain Deep Learning for Pairwise Identification of Strongly Lensed Gravitational-Wave Candidates}

\author[0009-0004-3509-7495]{Fan Zhang}
\email{fzhang@zju.edu.cn}  
\affiliation{%
 State Key Laboratory of Ocean Sensing {\normalfont\&} Ocean College, Zhejiang University, Zhoushan, 316021, China
}%
\affiliation{%
 Kavli Institute for Astrophysics and Space Research, Massachusetts Institute of Technology, Cambridge 02139, MA, USA
}%

\author[orcid=0009-0008-7902-8191]{Qikai Zhang}%
\email{15895935591@163.edu.cn}  
\affiliation{%
 State Key Laboratory of Ocean Sensing {\normalfont\&} Ocean College, Zhejiang University, Zhoushan, 316021, China
}%

\author[orcid=0009-0007-6444-2026]{Qiyuan Yang}
\email{yangqiyuan@whu.edu.cn}  
\affiliation{
 School of Physics and Technology, Wuhan University, Wuhan 430072, China
}%

\author[0009-0009-2516-8866]{Jiaqing Huang}
\email{moonspring233@gmail.com}
\affiliation{%
Jiangsu Sungo Intelligent Technology Research Center Co., Ltd., Jiangsu 226019, China
}%


\author[orcid=0009-0002-9540-9230]{Yong Yuan}
\email{yuanyong@imech.ac.cn}
\affiliation{Center for Gravitational Wave Experiment, National Microgravity Laboratory, Institute of Mechanics, Chinese Academy of Sciences, Beijing, China}

\author[orcid=0000-0002-8174-0128]{Xilong Fan}
\email[show]{Contact author: xilong.fan@whu.edu.cn}
\affiliation{
 School of Physics and Technology, Wuhan University, Wuhan 430072, China
}%
\correspondingauthor{Xilong Fan}

\date{\today}

\begin{abstract}
As gravitational wave (GW) catalogs continue to expand, exhaustive Bayesian comparisons of candidate event pairs become increasingly computationally expensive, which motivates the development of fast prescreening methods for strongly lensed GW searches. We formulate lensed-pair identification as a binary verification problem using two preprocessed strain segments. To address this task, we propose Physics-Inspired ResNet (PI-ResNet), a Siamese one-dimensional residual network for pairwise GW candidate classification. Unlike spectrogram-based prescreening approaches, PI-ResNet operates directly on whitened time-domain strain data and avoids an intermediate time--frequency image representation. A shared residual backbone with Squeeze-and-Excitation (SE) modules encodes the two input segments, and the paired embeddings are compared through absolute feature differences and Hadamard-product interactions. We train and evaluate the model using simulated GW signals from binary black hole mergers lensed by point-mass (PM) and singular isothermal sphere (SIS) lenses, injected into simulated LIGO and Einstein Telescope (ET) detector noise.
Under ET design noise, PI-ResNet achieves accuracies of $95.60\%$ for SIS lenses and $93.80\%$ for PM lenses, while maintaining $84.03\%$ and $78.25\%$ accuracy under simulated LIGO H1--L1 Gaussian noise. These results suggest that direct learning from 1D strain data provides an efficient and physically motivated preselection statistic for candidate lensed GW pairs, while also indicating the need for detector-domain adaptation.

\end{abstract}

\section{Introduction}
\label{sec:intro}

The Advanced LIGO detectors \cite{aasi2015advanced} enabled the first direct detection of GWs from a binary black hole (BBH) coalescence, GW150914, which marked the dawn of GW astronomy \cite{abbott2016observation}.
Since then, the LIGO-Virgo-KAGRA (LVK) collaboration has reported numerous GW events \cite{abbott2019gwtc, abbott2023search}.
As the sensitivity of ground-based GW detectors continues to improve toward the next-generation (3G) observatories, such as the ET \cite{punturo2010einstein} and Cosmic Explorer (CE) \cite{reitze2019cosmic}, the expected number of GW detections will increase by orders of magnitude.
This substantial increase in event rates will significantly enhance the probability of observing rare astrophysical phenomena, including the strong gravitational lensing of GWs \cite{Liao2017, Li2018}.
Strong gravitational lensing occurs when the GW signal emitted by a distant source travels through the spacetime curvature of a massive foreground object before reaching the observer \cite{ehlers2005gravitational}.
For macro-lenses like galaxy clusters, the time delay $\Delta t$ between multiple images typically ranges from days to months, resulting in distinct, repeated events in the detector strain \cite{sereno2010strong, baker2017multimessenger}.
In the geometrical-optics regime, strong lensing produces multiple GW images of the same compact-binary source. These images preserve the same intrinsic source parameters, but differ in their magnifications, arrival times, and
lensing-induced Morse phases \cite{Takahashi2003, dai2018detecting}. For macroscopic lenses, the multiple images may appear as separate candidate events in a GW catalog rather than as a single overlapping waveform.
Identifying whether two detected events are lensed counterparts of the same source is therefore a pairwise association problem.

Establishing such pairwise lensing associations in realistic detector noise presents a substantial computational and analytical challenge. Currently, matched filtering \cite{usman2016pycbc, messick2017analysis} remains the mathematically optimal approach. However, performing a coherent search across the vast, multidimensional parameter space of lens models and time delays is computationally prohibitive, especially when signals approach the physical detection limit (low SNR) in the ET era \cite{haris2018identifying,Hannuksela2019,pang2020lensed}. While deep learning has recently offered high-efficiency alternatives
\cite{george2018deep}, conventional convolutional neural networks (CNNs)
\cite{Simonyan2014VeryDC} and time--frequency image models often rely on
generic data representations. These representations do not explicitly encode the pairwise consistency
relations expected between multiple lensed images, including shared waveform
morphology, magnification differences, arrival-time shifts, and
lensing-induced phase-evolution differences. This motivates a model architecture that
operates on candidate pairs and incorporates lensing-inspired feature
interactions. 

Recent machine-learning studies have explored fast prescreening strategies for lensed-GW identification. \citet{goyal2021rapid} used Q-transform time--frequency maps and Bayestar sky-localization maps to rapidly filter candidate lensed pairs before Bayesian follow-up. \citet{magare2024slick} developed the SLICK pipeline by combining Q-transform maps with Sine-Gaussian projection maps to improve false-positive rejection. More recently, \citet{Li_2026} proposed SEMD, which formulates lensed-GW identification as a morphology-based similarity test on paired time--frequency spectrograms.

These studies demonstrate the potential of neural prescreening for large GW catalogs. However, most existing approaches rely on two-dimensional time--frequency or auxiliary image representations, which require an intermediate transformation of the detector strain. The consistency relations in amplitude and phase expected between multiple lensed images are also usually learned implicitly rather than encoded through explicit pairwise feature interactions. Among these methods, SEMD is the closest to our setting because it also formulates lensed-GW identification as a paired-representation classification problem. We therefore compare PI-ResNet with SEMD as the main image-based baseline.

In this study, we do not attempt a full catalog-level search over arbitrary event times, sky locations, or posterior samples. Instead, following the candidate-screening philosophy of recent machine-learning approaches, we formulate the problem as a binary pairwise verification task. Given two pre-triggered and preprocessed strain segments, the model predicts whether the pair is consistent with two lensed images of the same compact-binary source or with two unrelated events.

Within this setting, we propose PI-ResNet for rapid identification of candidate lensed GW pairs. The Physics-Fusion module compares paired latent features using their absolute difference and Hadamard product. This design provides a compact representation of waveform-morphology consistency and phase-coherence-related interactions between candidate images.


The remainder of this paper is organized as follows. Section~\ref{sec:lensed} reviews the GW lensing framework and the thin-lens models used in this work. Section~\ref{sec:Methods} describes the data generation, preprocessing, network architecture, and training procedure. Section~\ref{sec:Results} presents the classification results, baseline comparisons, and diagnostic analyses. Section~\ref{sec:Discussion} discusses the physical interpretation, error sources, limitations, and future outlook. Finally, Section~\ref{sec:Conclusion} summarizes the main findings. Throughout this paper, we take the speed of light $c=2.998\times 10^8 \mathrm{m}\cdot\mathrm{s}^{-1}$, the gravitational constant $G=6.67\times 10^{-11}\mathrm{m^3}\cdot\mathrm{kg}^{-1}\cdot\mathrm{s^{-2}}$ and the $\Lambda$CDM cosmology model described in \citet{2020A&A...641A...6P}, which is also coded in \textbf{Astropy}\footnote{http://www.astropy.org} \citep{2022ApJ...935..167A} with assumptions of  $H_{0}=67.66\mathrm{km}\cdot\mathrm{s}^{-1}\cdot\mathrm{Mpc}^{-1}$ and $\Omega_{\mathrm{m}}=0.30966$.


\section{Strongly Lensed GW Framework}\label{sec:lensed}
We briefly review the lensing of GWs in the regime relevant to ground-based detectors and then specialize to the PM and SIS models, following the scope and notation of prior work \citep{Takahashi2003,Kim2021, pagano2020lensinggw}.

\subsection{Gravitational lensing}\label{subsec:geo}
Depending on the GW wavelength relative to the characteristic lens scale,
one distinguishes between geometrical optics and wave optics approximations
\citep{ohanian1974focusing, Deguchi1986, Takahashi2003}. A convenient
criterion for the onset of diffraction is
\begin{equation}\label{eq:TNcriterion}
M_L \lesssim 10^{8}\,M_{\odot}\,\bigl(f/\mathrm{mHz}\bigr)^{-1},
\end{equation}
as discussed by \citet{Takahashi2003}. For the frequency band relevant to
ground-based compact-binary observations, $f\sim10^{2}$--$10^{3}\,\mathrm{Hz}$,
Equation~\eqref{eq:TNcriterion} indicates that lenses with masses around
$10^{2}$--$10^{3}\,M_{\odot}$ may enter the wave--optics regime, where
diffraction effects can no longer be neglected. Such low-mass micro- or
millilensing systems require wave-optics amplification factors and are not
the focus of the present work.

Instead, the lensing framework adopted below is the geometrical-optics
description, in which the lensed GW signal is represented as a superposition
of multiple images characterized by their magnifications, relative
arrival-time delays, and Morse phases. 

Under the thin–lens approximation the mass distribution is confined to a plane. Let $\gamma$ be the transverse source–plane displacement from the line of sight and $\xi$ the impact parameter on the lens plane. Using the Einstein radius $\xi_0$ as the length scale, the dimensionless source position is
\begin{equation}\label{eq:y_def}
y=\frac{\gamma\,D_L}{\xi_0\,D_S},
\end{equation}
where
\begin{equation}\label{eq:xi0_def}
\xi_0=\sqrt{\frac{4GM_L}{c^2}\,\frac{D_{LS}D_L}{D_S}} .
\end{equation}
Here $D_L$, $D_S$, and $D_{LS}$ are the angular–diameter distances from the observer to the lens, to the source, and from the lens to the source, respectively.

\subsubsection{Point-Mass Lens}\label{subsec:PM}
For a PM lens in the geometrical optics limit, the total amplification factor can be written in terms of the signed image magnifications $\mu_{\pm}$ and the inter–image time delay $\Delta t_d$ \cite{Takahashi2003, Kim2021}:
\begin{equation}\label{eq:F_PM}
F(f)= |\mu_{+}|^{1/2}-i\,|\mu_{-}|^{1/2}\,e^{-2\pi i f\,\Delta t_{d}}.
\end{equation}
We note that the sign in the exponential term differs from that used in \cite{Takahashi2003} because we adopt a different convention for the inverse Fourier transform.
The magnifications of the two images are given by:
\begin{equation}\label{eq:mu_PM}
\mu_{\pm} = \frac{1}{2} \pm \frac{y^2 + 2}{2y\sqrt{y^2 + 4}},
\end{equation}
and the relative arrival time delay between the two images is:
\begin{equation}\label{eq:dt_PM}
\Delta t_{d} = \frac{4\,G M_L^{z}}{c^{3}} \left[ \frac{y\sqrt{y^2+4}}{2} + \ln \left( \frac{\sqrt{y^2+4}+y}{\sqrt{y^2+4}-y} \right) \right],
\end{equation}
where $M_L^{z}\equiv M_L(1+z_L)$.
These equations completely specify the frequency-domain modulation induced by a PM lens, forming a two-image system for any $y>0$.

\subsubsection{Singular Isothermal Sphere Lens}\label{subsec:SIS}

For a SIS lens, the surface mass density is
characterized by the velocity dispersion $v$ as
$\Sigma(\xi)=v^2/(2G\xi)$, where $\xi$ is the impact parameter on the lens
plane. In the geometrical-optics limit, the total amplification factor can be
written in terms of the signed image magnifications $\mu_{\pm}$ and the
inter-image time delay $\Delta t_d$ \citep{Takahashi2003}:
\begin{equation}\label{eq:F_SIS}
F(f)=
\begin{cases}
|\mu_{+}|^{1/2}-i\,|\mu_{-}|^{1/2}\,e^{-2\pi i f\,\Delta t_{d}}, & y<1,\\[2pt]
|\mu_{+}|^{1/2}, & y\ge1.
\end{cases}
\end{equation}
The magnifications and the relative arrival time are
\begin{equation}\label{eq:mu_SIS}
\mu_{\pm}=1\pm\frac{1}{y},
\end{equation}
\begin{equation}\label{eq:dt_SIS}
\Delta t_{d}=8\,\frac{G M_L^{z}}{c^{3}}\,y,
\end{equation}
where $M_L^{z}$ denotes the redshifted mass enclosed within the Einstein
radius of the SIS lens. Equivalently, for an SIS lens,
\begin{equation}\label{eq:MLz_SIS}
M_L^{z}
=
\frac{4\pi^2 v^4}{G c^2}
\frac{D_LD_{LS}}{D_S}(1+z_L).
\end{equation}
When $y<1$, two images contribute and the second term in
Equation~\eqref{eq:F_SIS} produces the characteristic frequency-dependent
modulation. When $y\ge1$, a single image remains and $F(f)$ reduces
to a real magnification factor. Equations~\eqref{eq:F_SIS}--\eqref{eq:MLz_SIS}
fully specify the SIS prescription used in our simulations.


\begin{figure*}[ht!]
\centering
\includegraphics[width=\textwidth]{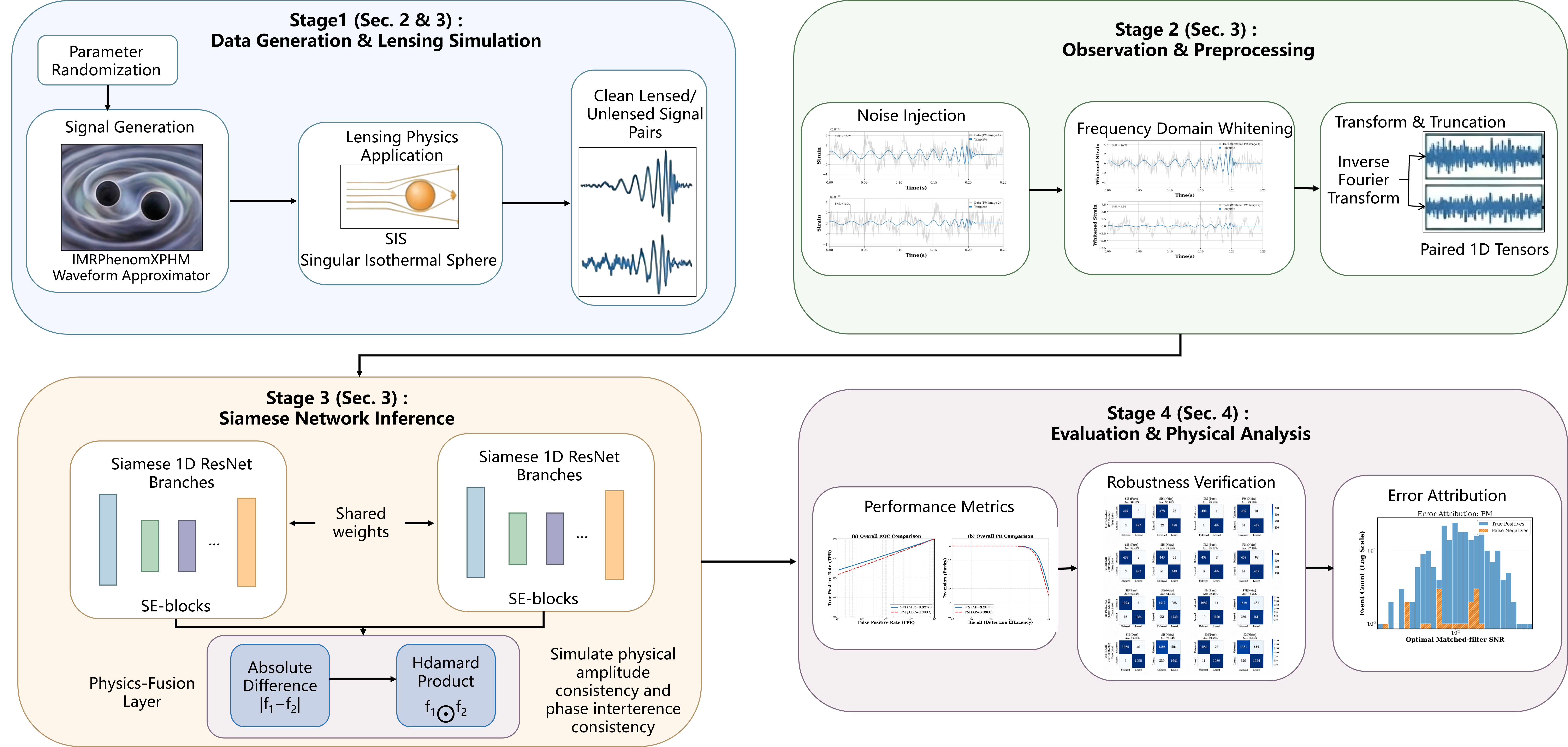} 
\caption{Workflow used for the simulated lensed-GW pairwise verification task. The pipeline includes waveform and lensing simulation (Stage 1), noise injection and whitening (Stage 2), Siamese 1D feature extraction and pairwise fusion (Stage 3), and performance evaluation (Stage 4).}
\label{fig:pipeline}
\end{figure*}

\subsection{Lensed GW Waveform}

In the frequency domain the lensed GW signal is related to the unlensed signal by
\begin{equation}\label{eq:hl_Fh}
h_L(f)=F(f)\,h(f),
\end{equation}
where $h(f)$ denotes the Fourier transform of the unlensed strain, $h_L(f)$ is the corresponding lensed strain, and $F(f)$ is the amplification factor specified by the lens model and the dimensionless source position $y$. In this work, we consider only lensing configurations that produce two images. We assume the time delay of the first image $\Delta t_{d1} = 0$, and it is the minimum image of the lens with Morse-phase parameter $\Delta\phi_1 = 0$. The second image is the saddle-point image with Morse-phase parameter $\Delta\phi_2 = 1/2$. The lensed GW waveforms $h_L^1(f)$ and $h_L^2(f)$ of the two images are then:
	\begin{equation}\label{eq.12}
		h^1_L(f) = \sqrt{|\mu_+|} \cdot h(f) \,,
	\end{equation}
	\begin{equation}\label{eq.13}
		h^2_L(f) = \sqrt{|\mu_-|} \cdot e^{-i \pi \left(2 f \Delta t_d + 1/2  \right)}\cdot h(f) \,.
	\end{equation}

To quantify the signal strength, we use the optimal signal-to-noise ratio (SNR). For a detector response $\tilde{h}(f)$ and the corresponding one-sided noise power spectral density (PSD) $S_n(f)$, the noise-weighted inner product is defined as
\begin{equation}
(a|b) = 4\,\mathrm{Re}\int_{f_{\min}}^{f_{\max}}
\frac{\tilde{a}^{*}(f)\tilde{b}(f)}{S_n(f)}\,df ,
\end{equation}
where $\tilde{a}(f)$ and $\tilde{b}(f)$ are frequency-domain waveforms, the asterisk denotes complex conjugation, and $S_n(f)$ is the one-sided detector noise PSD. Here, $f_{\min}$ and $f_{\max}$ denote the lower and upper frequency bounds used in the waveform generation and detector response calculation. The optimal SNR of a signal $h$ is then given by
\begin{equation}
\rho_{\rm opt} = \sqrt{(h|h)}
= \left[
4\int_{f_{\min}}^{f_{\max}}
\frac{|\tilde{h}(f)|^2}{S_n(f)}\,df
\right]^{1/2}.
\end{equation}

This definition follows the standard matched-filter formalism for compact-binary GW signals \citep{PhysRevD.49.2658}. In our simulations, the lensed frequency-domain waveform is first projected onto the detector response, and the optimal SNR is then computed from this response using the detector PSD. Because our ET simulations use a single detector, the reported SNR corresponds to a single-detector optimal SNR rather than a network SNR.

\section{Methods} \label{sec:Methods}

Figure~\ref{fig:pipeline} summarizes the workflow used in this study. We first simulate lensed and unlensed waveform pairs from randomized source and lens parameters. We then inject detector Gaussian noise for the ET and simulated LIGO-like configurations, whiten the detector strain in the frequency domain, and transform the data back to the time domain. The paired 1D strain inputs are passed to PI-ResNet, which outputs a matching probability. Finally, we evaluate the classification metrics and the SNR-dependent behavior of the model.

\subsection{Data Generation}
\label{subsec:data-prep}

\begin{deluxetable}{ll}
\tablecaption{Parameters and distributions used for simulated source and lens populations.\label{tab:source-lens-params}}
\tablehead{
\colhead{Parameter} & \colhead{Range / distribution}
}
\startdata
\multicolumn{2}{@{}l}{\textbf{Source parameters}} \\
Primary and secondary source-frame masses $m_1,m_2$ & Uniform $[10,70]\,M_\odot$ \\
Spin magnitudes $a_1,a_2$ & Uniform $[0,0.99]$ \\
Spin tilt angles & Sine prior on $[0,\pi]$ \\
Azimuthal spin angles $\phi_{12},\phi_{jl}$ & Uniform $[0,2\pi)$ \\
Inclination $\theta_{jn}$ & Sine prior on $[0,\pi]$ \\
Polarization $\psi$ & Uniform $[0,\pi)$ \\
Coalescence phase $\phi_c$ & Uniform $[0,2\pi)$ \\
Sky location $(\alpha,\delta)$ & RA uniform $[0,2\pi)$ and Dec cosine prior \\
Source redshift $z_s$ & $[0.01,1]$, calculated by $d_L$ with a uniform prior in comoving volume \\
\hline
\multicolumn{2}{@{}l}{\textbf{PM lens parameters}} \\
Lens redshift $z_L$ & $z_L=z_s/2$ \\
Point-lens mass $M_L$ & Uniform $[10^8,10^{10}]\,M_\odot$ \\
Impact parameter $y$ & Uniform $[0.01,0.3]$ \\
\hline
\multicolumn{2}{@{}l}{\textbf{SIS lens parameters}} \\
Lens redshift $z_L$ & $z_L=z_s/2$ \\
Velocity dispersion $\sigma_v$ & Uniform $[100,500]\,\mathrm{km\,s^{-1}}$ \\
Impact parameter $y$ & Uniform $[0.01,0.3]$ \\
\enddata
\end{deluxetable}

\begin{table*}[t]
\centering
\caption{Fixed simulation settings}
\label{tab:sim-settings}
\begin{tabular}{l l}
\hline\hline
Setting & Value \\
\hline
Waveform approximant & \texttt{IMRPhenomXPHM} \\
Minimum and reference frequency & $f_{\min}=20~\mathrm{Hz}$ and $f_{\mathrm{ref}}=10~\mathrm{Hz}$ \\
Sampling rate & $4096~\mathrm{Hz}$ \\
Segment length & $24~\mathrm{s}$ \\
ET detector setting & Single ET-D interferometer with PSD in \texttt{Bilby} \\
LIGO detector setting & H1--L1 with PSDs in \texttt{Bilby} \\
Whitening & Frequency-domain whitening followed by inverse FFT to the time domain \\
\hline
\end{tabular}
\end{table*}

We first generate source-level lensed and unlensed strain segments before constructing candidate pairs. The underlying BBH waveforms are generated with \texttt{IMRPhenomXPHM} \citep{2019PhRvD.100b4059K, 2021PhRvD.103j4056P} through the \texttt{Bilby}\footnote{https://git.ligo.org/lscsoft/bilby}~\citep{2019ApJS..241...27A} interface and converted into $24~\mathrm{s}$ time-domain segments sampled at $4096~\mathrm{Hz}$, corresponding to $N=98304$ samples. The source and lens parameters are summarized in Table~\ref{tab:source-lens-params}, and the fixed simulation settings are listed in Table~\ref{tab:sim-settings}. The PM mass range and the SIS velocity-dispersion range are used to define macrolensing, geometrical-optics simulations in which multiple images are treated as separated candidate events.

For lensed systems, the two images of the same compact-binary source are generated from the same source-lens configuration and stored as separate pre-triggered event segments. They share the same intrinsic source parameters but differ in magnification, arrival time, and image-type phase factors. The inter-image delay is therefore not required to fit within a single $24~\mathrm{s}$ data segment, and the network does not use a summed multi-image waveform as input. The SIS samples used in this work are restricted to $0.01 \leq y \leq 0.3$, corresponding to the two-image regime. The construction of positive and negative candidate pairs from these source-level segments is described in Section~\ref{subsec:training_details}.

For each lens model, we generate both noise-free and noise-injected datasets. For clarity, we refer to the noise-free datasets as ``Pure'' and the Gaussian-noise-injected datasets as ``Noisy'' in the following tables and figures. In the noisy case, the simulated signals are projected onto the detector response, injected into Gaussian noise generated from the target detector PSD, whitened in the frequency domain, and transformed back to the time domain.The resulting whitened 1D strain segments are used directly as inputs to PI-ResNet, without constructing time--frequency images.

\begin{figure}[t]
\centering
\includegraphics[width=\columnwidth]{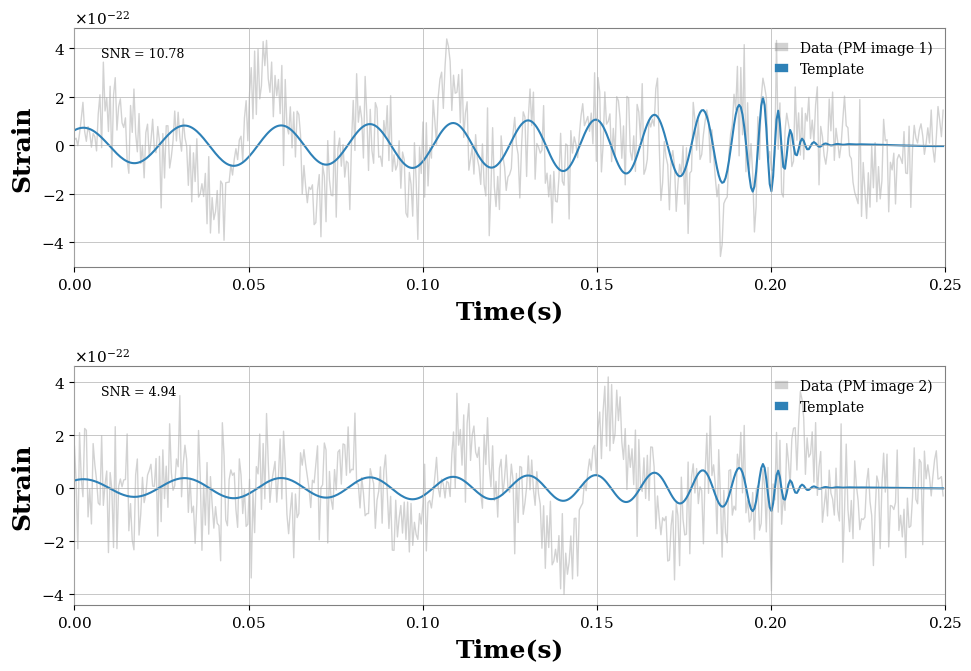}
\caption{Representative unwhitened time-domain strain segments for the two images of the same PM-lensed GW event. The gray curves denote the noisy detector strains, and the blue curves denote the corresponding pure templates. The two images show similar chirp morphology but different amplitudes and SNRs, reflecting the different magnifications of the lensed images.}
\label{fig:lensed_pair_raw}
\end{figure}

Figure~\ref{fig:lensed_pair_raw} shows a representative pair of unwhitened time-domain strain segments corresponding to the two images of the same PM-lensed GW event. The gray curves denote the noisy detector strains, and the blue curves denote the corresponding pure templates. The two segments exhibit the same underlying chirp morphology but different amplitudes and SNRs, consistent with the different magnifications of the two lensed images. In the subsequent preprocessing step, these strain segments are whitened and then used as paired inputs to the classifier.

\begin{figure}[t]
\centering
\includegraphics[width=\columnwidth]{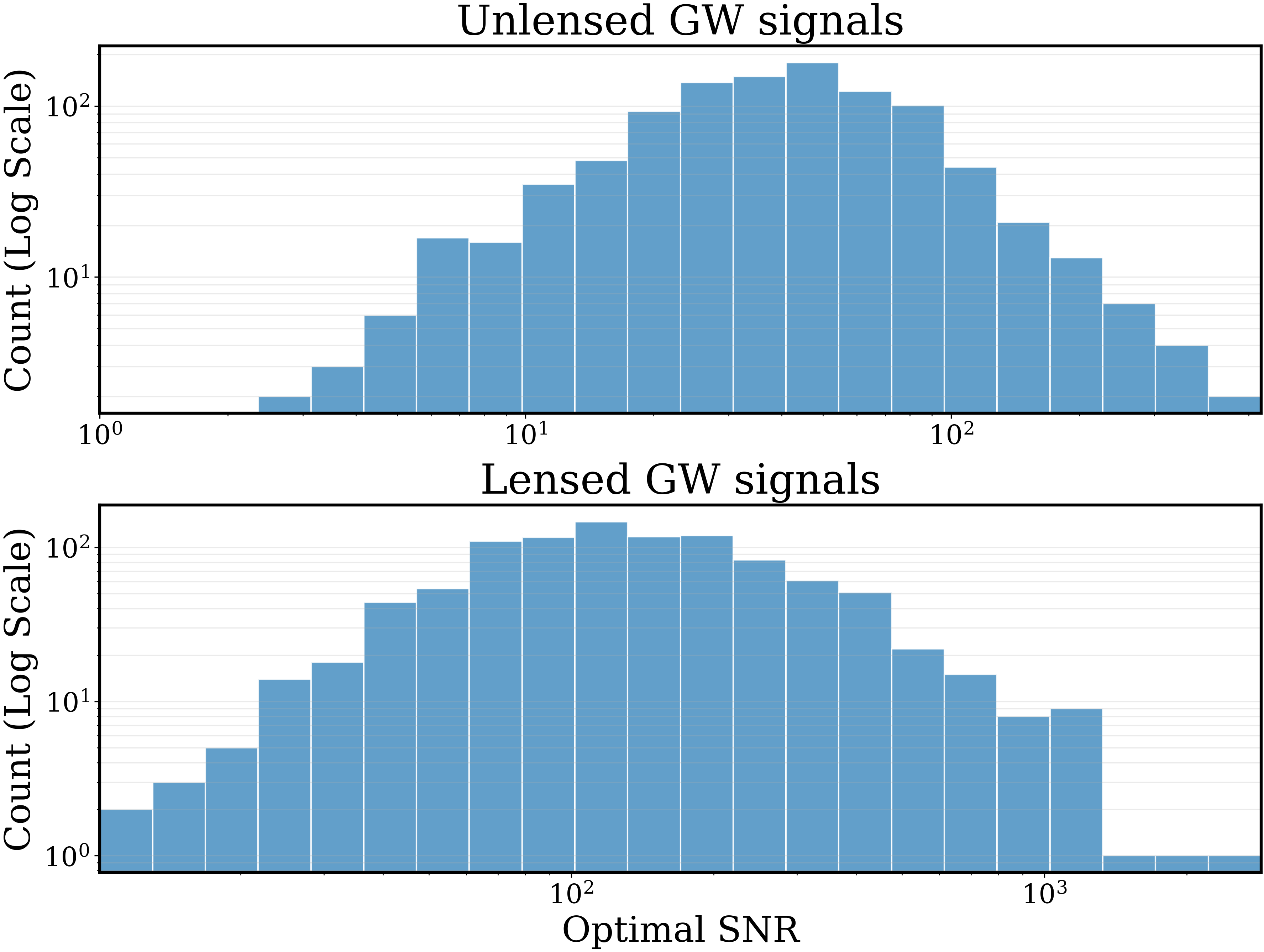}
\caption{Optimal SNR distributions of the simulated unlensed and lensed GW signals. 
The vertical axis is shown on a logarithmic scale. 
Lensing magnification shifts part of the lensed population toward higher SNR values, producing a broader high-SNR tail.}
\label{fig:snr_dist}
\end{figure}

Figure~\ref{fig:snr_dist} shows the optimal SNR distributions of the simulated unlensed and lensed signals. 
The lensed population extends to higher SNR values because lensing magnification increases the observed strain amplitude for part of the simulated events.

\subsection{Architecture of PI-ResNet}
\label{sec:siamese_arch}

\begin{figure*}[t!]
\centering
\includegraphics[width=0.9\textwidth]{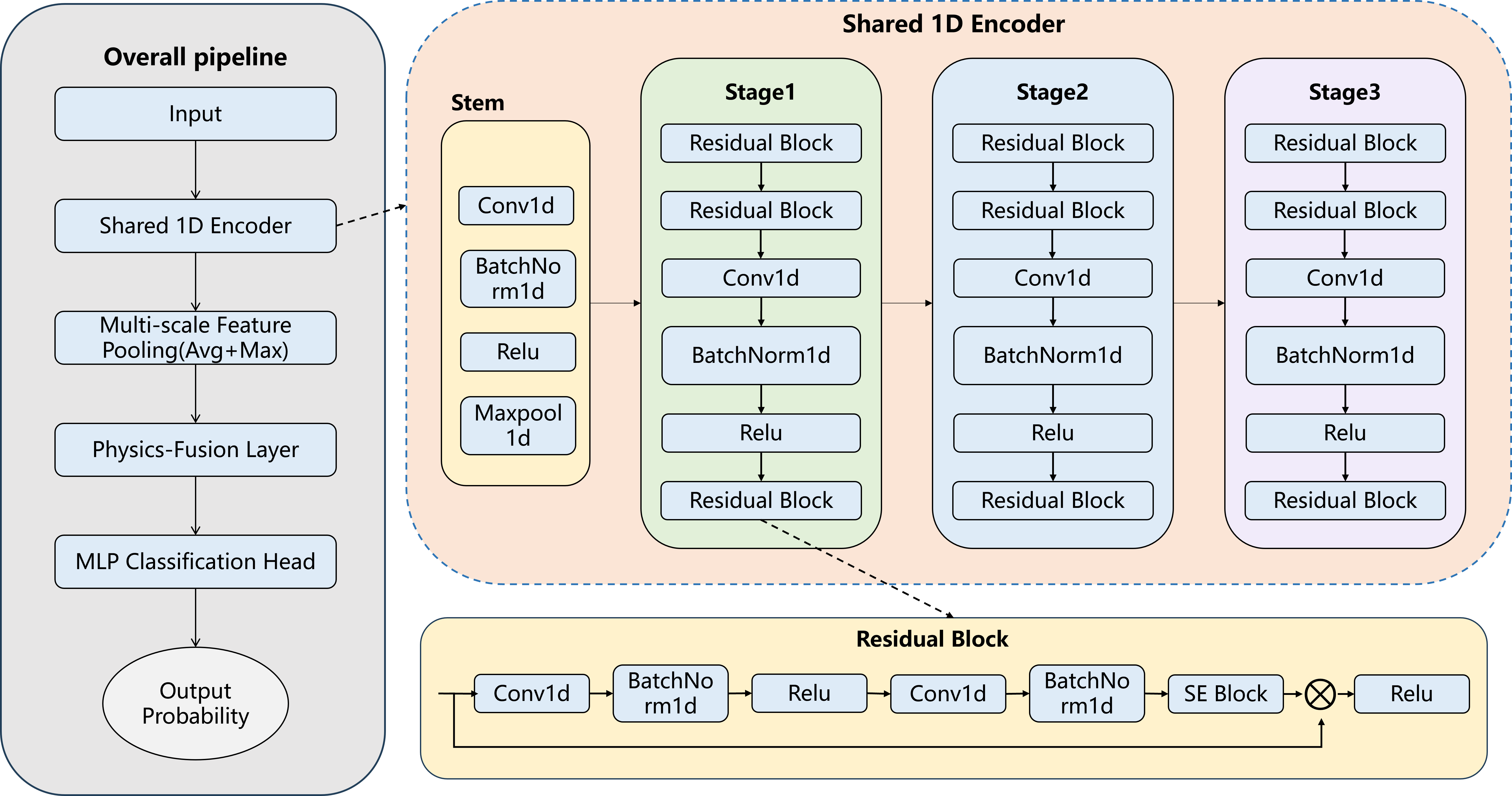}
\caption{End-to-end PI-ResNet architecture. \textbf{Left:} Two candidate whitened strain segments, denoted as L1 and L2, are processed by a shared Siamese backbone. \textbf{Right:} Enlarged view of the \texttt{ResBlock}, which integrates SE modules. The final output is classified based on the fused pairwise representation.}
\label{fig:architecture}
\end{figure*}

To determine whether two independent GW events are lensed counterparts, we
design an end-to-end PI-ResNet. Each input is a pre-triggered and whitened
strain segment that has been aligned according to the injection coalescence
time in our simulations. Before being passed to the network, the strain is
converted into a compact fixed-length one-dimensional representation through
a deterministic temporal standardization and decimation procedure. This step standardizes the temporal dimension of all inputs and reduces the computational cost while keeping the same preprocessing protocol for both branches of the Siamese network. The same procedure is applied to
both branches of the Siamese network, resulting in paired one-dimensional
input tensors of length \(T=4096\). Our \texttt{PI-ResNet} operates directly
on these processed pairs of time series. The high-level roadmap of this
architecture is illustrated in Figure~\ref{fig:architecture}.

\subsubsection{Shared Siamese Backbone and SE Attention}

The two input branches share identical weights to project the paired detector
strains into a common latent space \citep{bromley1993signature}. Each branch
uses a 1D residual convolutional backbone to extract local time-domain
features from the whitened strain. The shared-weight design ensures that the
two candidate segments are processed by the same feature extractor before
pairwise fusion.

We incorporate 1D SE modules into the residual
blocks \citep{hu2018squeeze}. These modules adaptively reweight channel-wise
features after convolutional extraction, allowing the backbone to emphasize
informative strain patterns while reducing the contribution of less relevant
channels. The resulting latent representations are then passed to the
Physics-Fusion layer for pairwise comparison.

\subsubsection{Physics-Fusion Mechanism}

Rather than using generic feature concatenation, our pairwise verification
module adopts a physically motivated Physics-Fusion layer. Let
\(\mathbf{f}_1\) and \(\mathbf{f}_2\) denote the learned representations of
the two input strains extracted by the shared Siamese backbone. The fused
pairwise representation is defined as
\begin{equation}
\mathbf{f}_{\mathrm{fusion}}
=
\left[
|\mathbf{f}_1-\mathbf{f}_2|,\;
\mathbf{f}_1\odot\mathbf{f}_2
\right],
\label{eq:physics_fusion}
\end{equation}
where \(\odot\) denotes the standard element-wise Hadamard product
\cite{chrysos2025hadamard}. This design is intended to represent two
lensing-relevant feature interactions for catalog-level candidate-pair
verification in the geometrical-optics multiple-image regime:
amplitude-envelope-related differences and phase-coherence-related
multiplicative interactions. It is also closely related to the pairwise
consistency tests used in traditional Bayesian lensing searches
\cite{janquart2021fast}.

The absolute-difference branch,
\(|\mathbf{f}_1-\mathbf{f}_2|\), measures residual discrepancies between
the two learned waveform representations. Since multiple lensed images
originate from the same compact-binary source, their normalized
macroscopic strain envelopes should remain consistent after accounting
for magnification. This branch therefore helps reject unrelated or
noise-dominated pairs with inconsistent waveform morphology.

The Hadamard-product branch, \(\mathbf{f}_1\odot\mathbf{f}_2\), provides a
multiplicative interaction between the two latent representations. Its
motivation comes from the geometrical-optics expression for a lensed image
\cite{Abbott_2021},
\begin{equation}
\tilde{h}^{L}_{j}(f)
=
\sqrt{|\mu_j|}
\tilde{h}(f;\theta)
\exp\left[
-i\pi\left(
2 f\Delta t_j
+
\Delta\phi_j
\right)
\right],
\label{eq:lensed_image_go}
\end{equation}
where \(\theta\) denotes the intrinsic source parameters, \(\mu_j\) is the
magnification, \(\Delta t_j\) is the arrival-time delay, and
\(\Delta\phi_j\) is the Morse phase. Equation
\eqref{eq:lensed_image_go} shows that multiple images of the same source
share the same intrinsic waveform while differing mainly through
magnification, arrival time, and Morse phase.

Using Equation~\eqref{eq:lensed_image_go}, the frequency-domain cross term
for two candidate images from the same source can be written as
\begin{equation}
\begin{aligned}
\tilde{h}^{L*}_{1}(f)\tilde{h}^{L}_{2}(f)
&=
\sqrt{|\mu_1\mu_2|}
|\tilde{h}(f;\theta)|^2  \\
&\quad \times
\exp\left[
-i \pi\left(
 2f\Delta t_{21}
+
\Delta\phi_{21}
\right)
\right],
\end{aligned}
\label{eq:lensed_cross_term}
\end{equation}
where \(\Delta t_{21}=\Delta t_2-\Delta t_1\) and
\(\Delta\phi_{21}=\Delta\phi_2-\Delta\phi_1\). This expression motivates the use of multiplicative feature interactions,
since such interactions can emphasize latent components that remain
correlated between candidate images despite lensing-induced magnification
differences, arrival-time offsets, and phase-evolution changes. We do not
claim that the Hadamard branch is an exact analytic substitute for a
Morse-phase test, nor do we isolate its response to Morse phase in this work.
Rather, it provides a compact neural feature-interaction mechanism for
suppressing incoherent alignments caused by unrelated events or detector
noise.

\subsection{Dataset Construction and Training Details}
\label{subsec:training_details}

To prevent source-level data leakage, we split the simulated data before pair construction. For the ET design-noise experiments, each lens-model catalog contains $2,500$ independent source-level indices, each corresponding to two lensed images generated from the same compact-binary source. For the simulated LIGO H1--L1 Gaussian-noise comparison, each lens-model catalog contains $10,000$ independent source-level indices. For each source index, one positive pair is formed by pairing the two lensed images of the same source. Negative pairs are constructed either by pairing images from different lensed sources or by pairing a lensed image with an unlensed background segment. This yields $5,000$ candidate pairs for each ET dataset, consisting of $2,500$ positive pairs and $2,500$ negative pairs. For the simulated LIGO H1--L1 datasets, this yields $20,000$ candidate pairs, consisting of $10,000$ positive pairs and $10,000$ negative pairs. We adopt a larger source catalog for the LIGO H1–L1 experiments because the higher-noise regime requires more training samples to converge.

For the ET datasets, the $2,500$ source-level indices are randomly split into $80\%$ for training and $20\%$ for validation. For the simulated LIGO H1--L1 datasets, the same source-level splitting strategy is applied to the larger paired catalog. In all cases, the split is performed before pair generation, so no physical source appears in both subsets. During training, negative pairs are dynamically sampled, with $70\%$ hard negatives from mismatched lensed sources and $30\%$ easy negatives from lensed--unlensed pairs. During validation, all positive and negative pairs are fixed using the same random seed for reproducibility.

The \texttt{PI-ResNet} is trained using the Binary Cross-Entropy with Logits
loss (\texttt{BCEWithLogitsLoss}). We optimize the model parameters with the
AdamW optimizer using an initial learning rate of $10^{-4}$ and a weight
decay of $10^{-4}$. The learning rate is annealed to $10^{-6}$ with a cosine
annealing schedule. The maximum number of training epochs is set to $300$,
and the pairwise verification threshold is fixed at $0.5$. The model
checkpoint with the highest validation accuracy is retained. Therefore, the
metrics reported under this protocol should be interpreted as validation-set
performance rather than independent test-set performance.

\section{Results} \label{sec:Results}

\subsection{Detection Sensitivity and Baseline Comparisons}
\label{subsec:performance}

We evaluate PI-ResNet using simulated BBH signals under the ET design-noise
setting and compare it with the 2D SEMD baseline \citep{Li_2026}. In SEMD,
the same 1D strain segments are first converted into Constant-Q Transform
(CQT) time--frequency maps before pairwise classification. In contrast,
PI-ResNet operates directly on whitened 1D strain pairs.

PI-ResNet and SEMD are evaluated with the same source-level splits, noise realizations, training budget, and maximum number of epochs. For SEMD, we additionally perform a grid search over CQT-related preprocessing parameters and training hyperparameters, including the time--frequency resolution, analyzed frequency range, and learning rate. 

\begin{table}[htbp]
\centering
\caption{Performance comparison between the proposed 1D PI-ResNet and the 2D image-based SEMD framework under the ET design-noise setting.}
\label{tab:comparison_semd}
\renewcommand{\arraystretch}{1.2}
\begin{tabular}{llcc}
\hline\hline
Task & Model & AUC & Accuracy \\
\hline
SIS-Pure  & \textbf{PI-ResNet} & \textbf{1.0000} & \textbf{99.40\%} \\
          & SEMD (2D) & 0.9993          & 98.40\% \\ \hline
SIS-Noisy & \textbf{PI-ResNet} & \textbf{0.9910} & \textbf{95.60\%} \\
          & SEMD (2D) & 0.9726          & 89.80\% \\ \hline
PM-Pure   & \textbf{PI-ResNet} & \textbf{1.0000} & \textbf{99.80\%} \\
          & SEMD (2D) & 0.9999          & 99.50\% \\ \hline
PM-Noisy  & \textbf{PI-ResNet} & \textbf{0.9897} & \textbf{93.80\%} \\
          & SEMD (2D) & 0.9500          & 87.70\% \\
\hline\hline
\end{tabular}
\end{table}

\begin{table}[htbp]
\centering
\caption{Performance comparison between the proposed 1D PI-ResNet and the 2D image-based SEMD framework under the simulated LIGO H1--L1 Gaussian-noise setting.}
\label{tab:comparison_ligo}
\renewcommand{\arraystretch}{1.2}
\begin{tabular}{llcc}
\hline\hline
Task & Model & AUC & Accuracy \\
\hline
SIS-Pure  & \textbf{PI-ResNet} & \textbf{0.9998} & \textbf{99.42\%} \\
          & SEMD (2D) & 0.9996          & 98.88\% \\ \hline
SIS-Noisy & \textbf{PI-ResNet} & \textbf{0.9168} & \textbf{84.03\%} \\
          & SEMD (2D) & 0.8762          & 78.43\% \\ \hline
PM-Pure   & \textbf{PI-ResNet} & \textbf{0.9998} & \textbf{99.48\%} \\
          & SEMD (2D) & 0.9996          & 99.23\% \\ \hline
PM-Noisy  & \textbf{PI-ResNet} & \textbf{0.8685} & \textbf{78.25\%} \\
          & SEMD (2D) & 0.8346          & 74.37\% \\
\hline\hline
\end{tabular}
\end{table}

\begin{figure*}[ht!]
    \centering
    \includegraphics[width=0.9\textwidth]{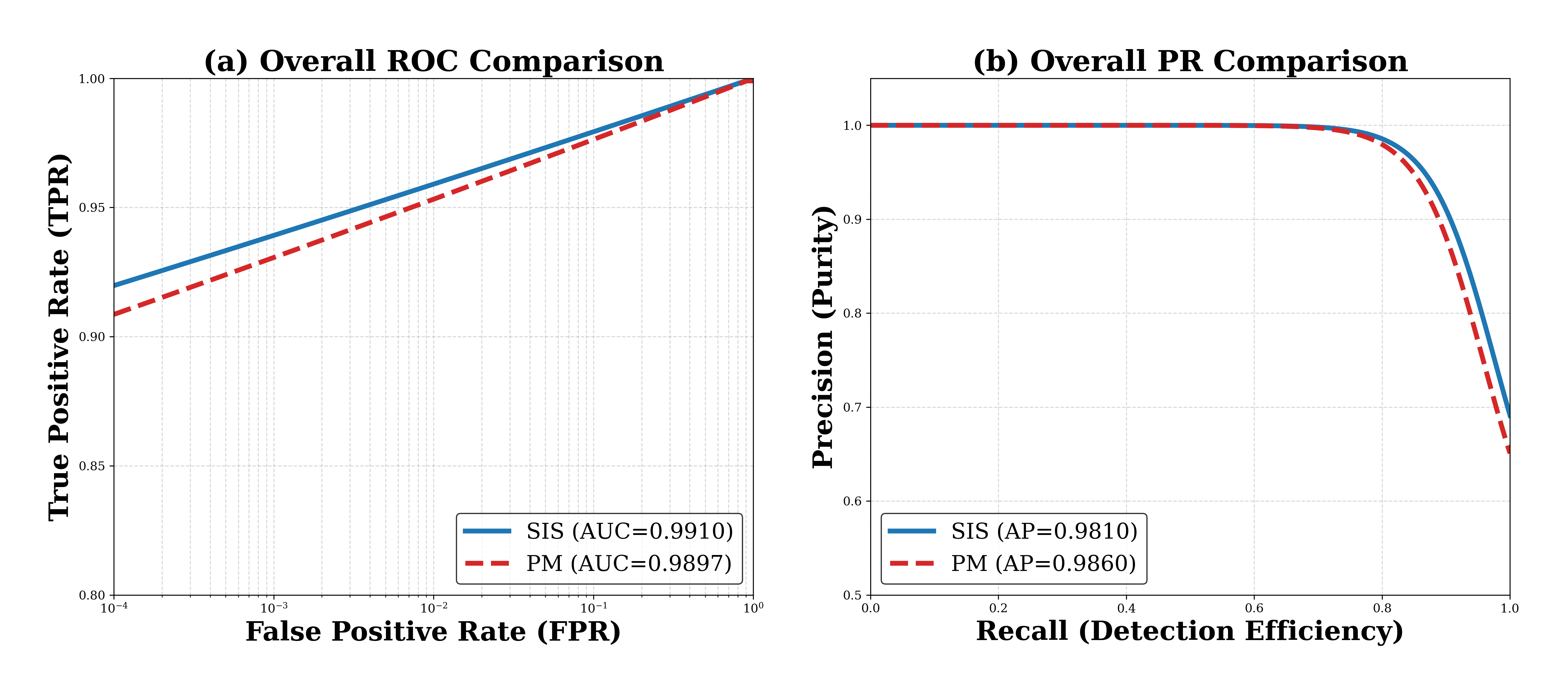}
    \caption{ROC and PR curves for SIS and PM lens models in the noisy ET setting. \textit{Left:} ROC curves, with the TPR axis restricted to $0.8$--$1.0$ for clarity. \textit{Right:} PR curves.}
    \label{fig:performance_noisy}
\end{figure*}

Table~\ref{tab:comparison_semd} compares PI-ResNet with SEMD under ET design-noise conditions. In the pure setting, both models achieve very high classification performance. After ET Gaussian noise is added, PI-ResNet maintains higher accuracy than SEMD, outperforming it by $5.80$ percentage points for SIS-Noisy and $6.10$ percentage points for PM-Noisy.

The same comparison is extended to simulated LIGO H1--L1 Gaussian-noise conditions in Table~\ref{tab:comparison_ligo}. Both models perform well in the pure control cases. Their performance decreases under simulated LIGO H1--L1 Gaussian noise, suggesting sensitivity to changes in the detector-noise PSD. PI-ResNet still outperforms SEMD, reaching accuracies of $84.03\%$ for SIS and $78.25\%$ for PM, corresponding to improvements of $5.60$ and $3.88$ percentage points, respectively.

Taken together, these results suggest that direct classification of whitened 1D strain pairs can provide a useful alternative to intermediate time--frequency image representations. The improvement is consistent across the simulated ET and LIGO settings. Since the LIGO dataset is generated from detector PSDs rather than real non-Gaussian strain data, the observed degradation should not be attributed to glitches. Instead, it mainly reflects the difficulty of transferring the classifier to a different detector-noise PSD. These results support the use of PI-ResNet as a lightweight preselection model, while also indicating that detector-domain adaptation will be needed for more robust deployment.

\subsection{Visualization of Classification Performance}
\label{subsec:classification_efficacy}

\begin{figure*}[htbp]
    \centering
    \includegraphics[width=\textwidth]{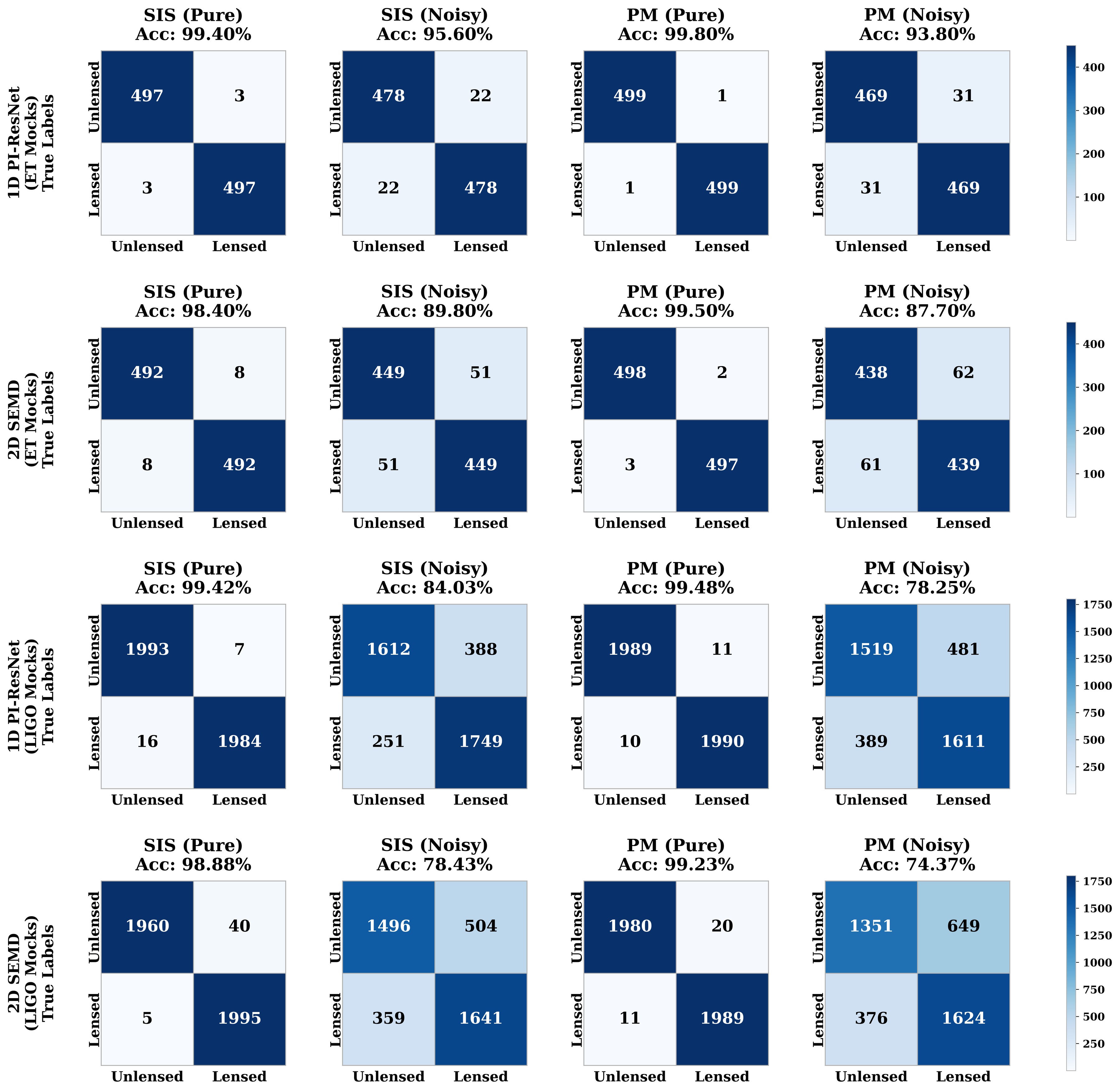}
    \caption{Confusion-matrix summary for pairwise classification across the
    main evaluation scenarios. \textbf{Top row:} 1D PI-ResNet evaluated on pure
    and ET-design Gaussian-noise mock data for the SIS and PM lens models.
    \textbf{Second row:} 2D SEMD baseline evaluated under the corresponding pure
    and ET-design Gaussian-noise mock configurations. \textbf{Third row:} 1D
    PI-ResNet evaluated on the LIGO-comparison datasets, including the
    pure controls and the simulated H1--L1 Gaussian-noise cases.
    \textbf{Fourth row:} 2D SEMD evaluated under the corresponding
    LIGO-comparison configurations. Each panel reports the validation-set
    accuracy and the confusion matrix for unlensed/lensed pairwise
    classification.}
    \label{fig:master_confusion_matrix}
\end{figure*}

To complement the tabulated results, we further visualize the classifier performance using ROC curves, PR curves, and confusion matrices. These plots provide a direct view of the model behavior under noisy evaluation settings.

Figure~\ref{fig:performance_noisy} shows the ROC and PR performance of PI-ResNet under ET-design Gaussian noise. The SIS and PM cases give similar ROC curves, with AUC values of $0.9910$ and $0.9897$, respectively. Their PR curves also remain stable, with average precision values of $0.9810$ for SIS and $0.9860$ for PM. These results indicate that PI-ResNet maintains strong discrimination for both lens models in the noisy ET validation set.

Figure~\ref{fig:master_confusion_matrix} confirms the trends observed in the quantitative comparisons. In the ET mock-data experiments, PI-ResNet shows stronger diagonal dominance than SEMD. Under ET design noise, PI-ResNet reaches $95.60\%$ accuracy for SIS and $93.80\%$ for PM, compared with $89.80\%$ and $87.70\%$ for SEMD. Under simulated LIGO H1--L1 Gaussian noise, both models show lower accuracy than in the ET setting. PI-ResNet still outperforms SEMD, reaching $84.03\%$ for SIS and $78.25\%$ for PM, compared with $78.43\%$ and $74.37\%$ for SEMD. These results suggest that the 1D strain-based model is less sensitive to the simulated detector-noise setting than the 2D time--frequency baseline, while detector-PSD domain shift remains an important limitation.



\subsection{SNR-Dependent Error Diagnostics}
\label{subsec:sensitivity_snr}

\begin{figure}[t]
\centering

\includegraphics[width=0.95\columnwidth]{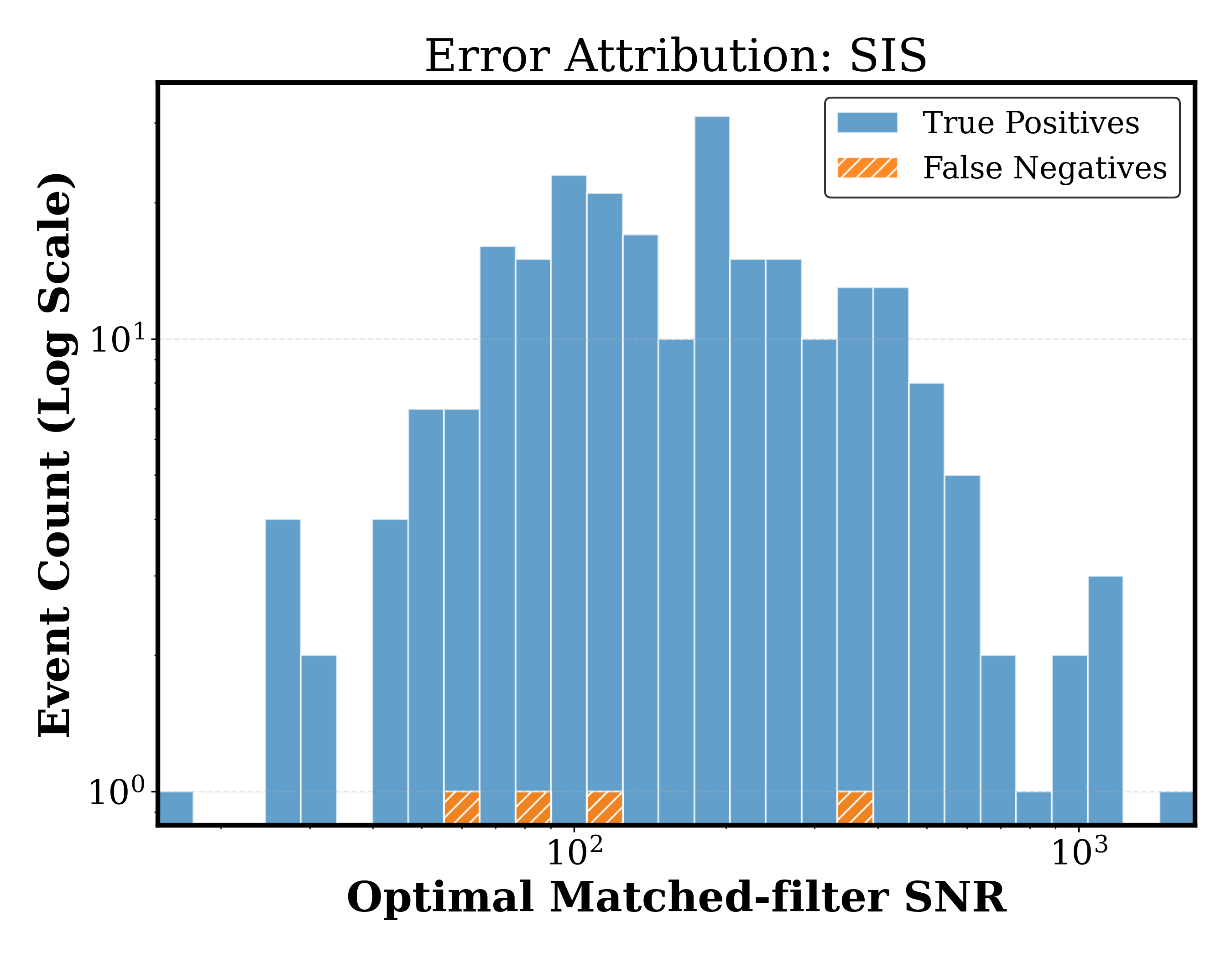}

\vspace{0.3em}
{\small (a) SIS Lens Model}

\vspace{0.8em}

\includegraphics[width=0.95\columnwidth]{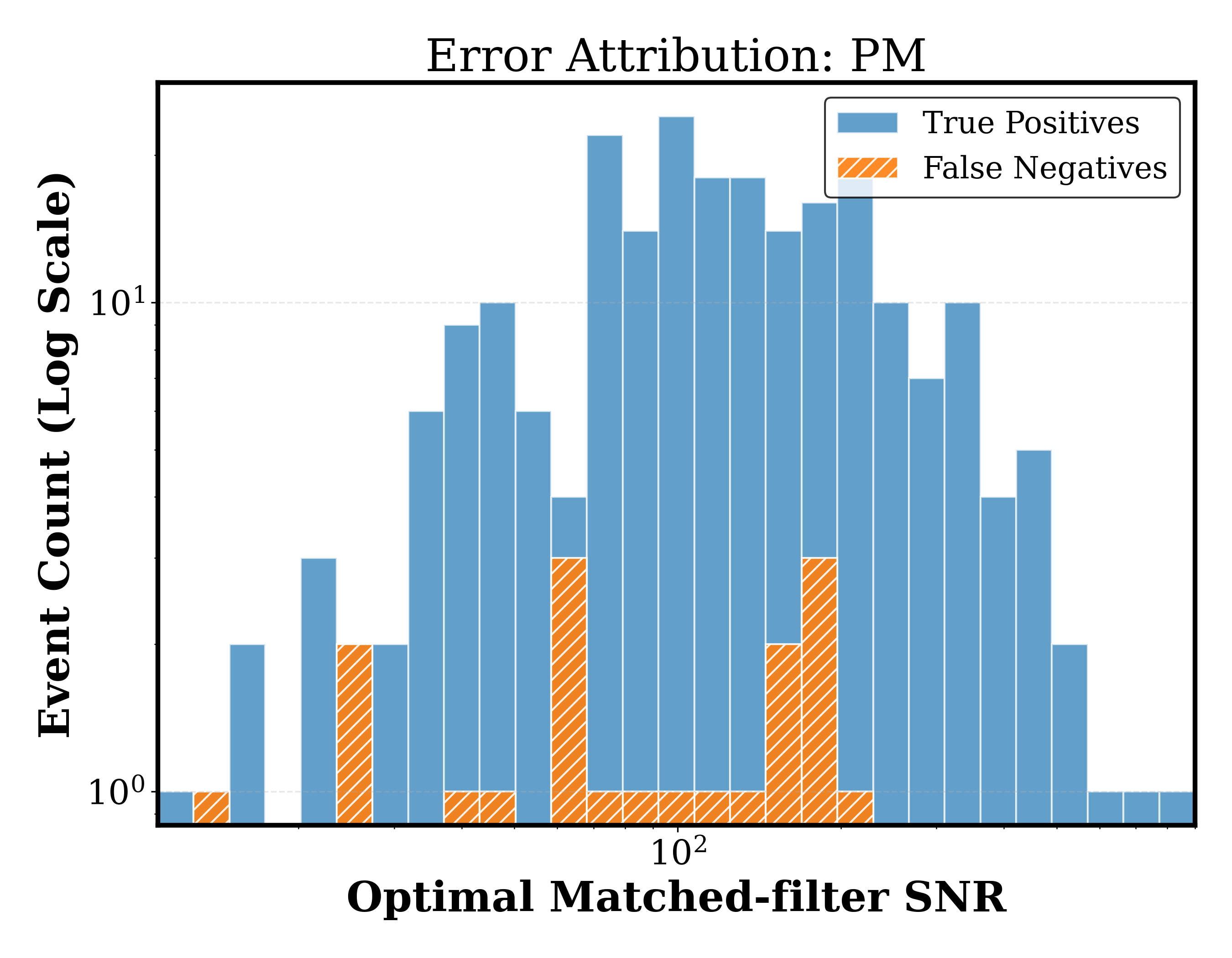}

\vspace{0.3em}
{\small (b) PM Lens Model}

\caption{Distribution of true positives and false negatives as a function of optimal matched-filter SNR for the noisy ET validation samples. The SNR axis is shown on a logarithmic scale. The histograms provide an error-attribution view of where missed associations occur within the sampled SNR range.}
\label{fig:error_by_snr_large}
\end{figure}

\begin{figure}[ht!]
\centering
\includegraphics[width=\columnwidth]{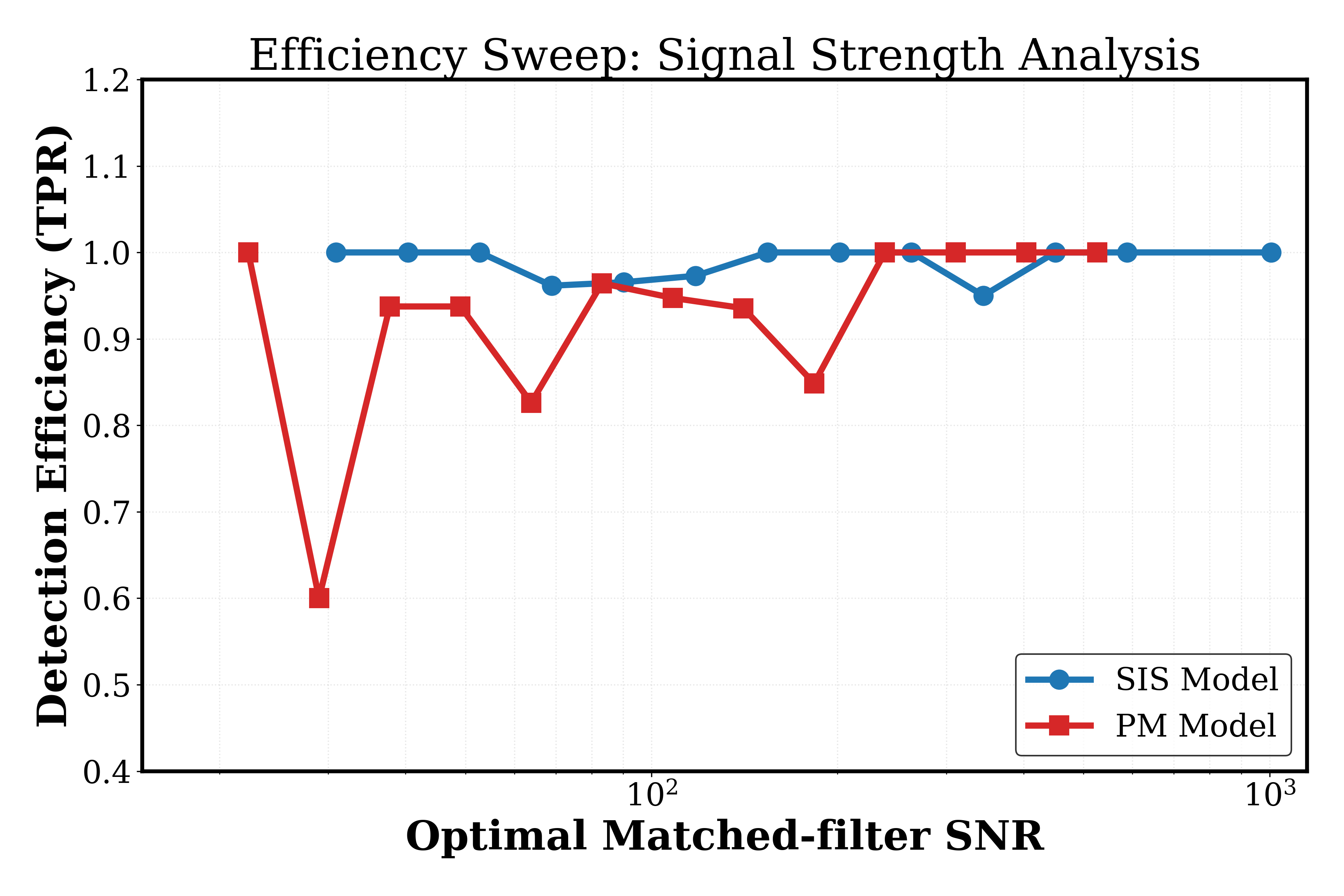}
\caption{Pairwise association efficiency, measured by the 
true-positive rate (TPR), as a function of optimal matched-filter SNR for the noisy ET validation samples. The TPR is computed in logarithmic SNR bins. The trend shows that the association efficiency remains high over most of the sampled range and tends to saturate at higher SNR. Statistical fluctuations in low-occupancy bins ($N < 10$ samples) appear as bin-to-bin variations rather than systematic trends.}
\label{fig:efficiency_sweep_combined}
\end{figure}

\begin{figure}[t]
\centering
\includegraphics[width=\columnwidth]{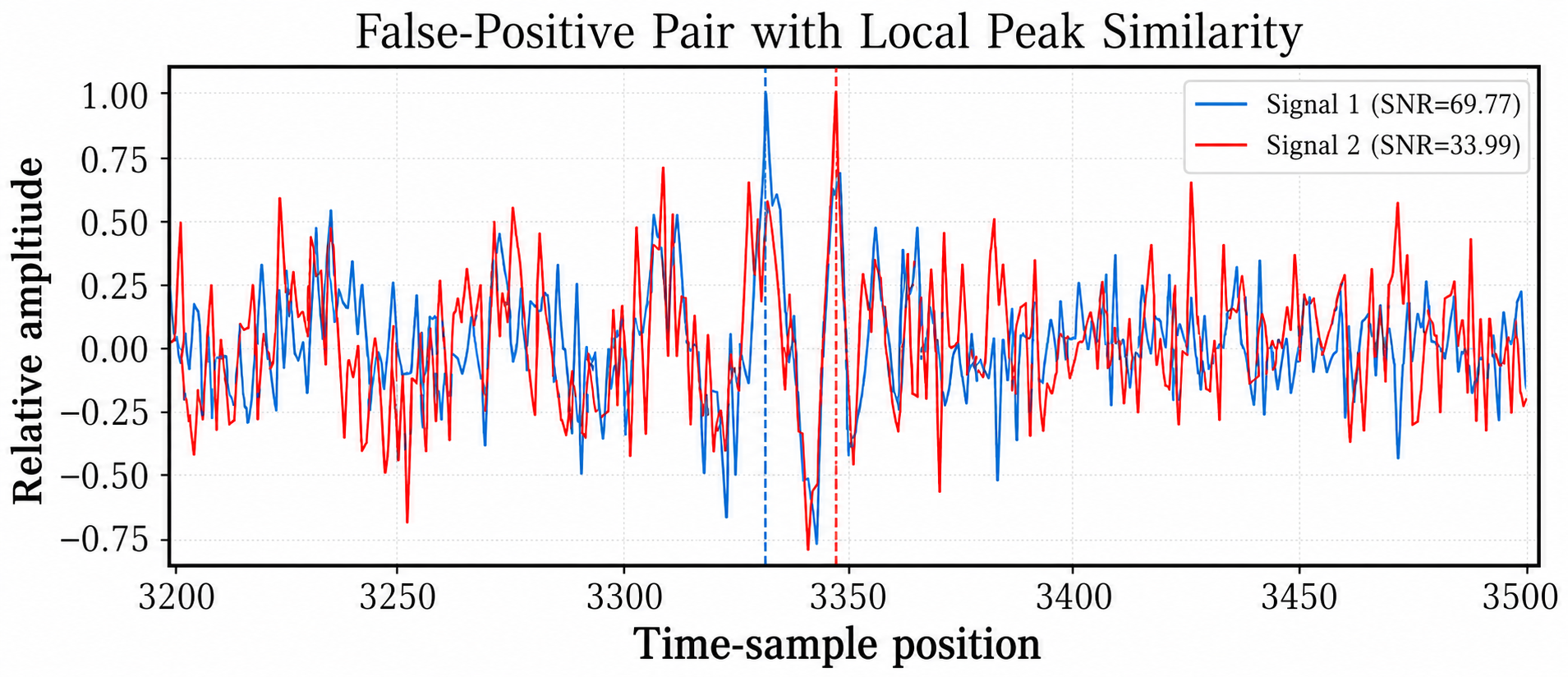}
\caption{Representative false-positive pair in the noisy ET validation set. 
The two signals are not generated from the same lensed source, but they show locally similar peak structures within a narrow time-sample region. 
The dashed vertical lines mark the dominant local peaks of the two signals. 
This example illustrates how unrelated pairs can produce short local waveform similarities that lead to spurious pairwise associations.}
\label{fig:false_positive_local}
\end{figure}

Signal strength affects lensed-pair verification, but it is not the only factor controlling the classification difficulty. 
We therefore group the noisy ET validation samples by their optimal matched-filter SNR and examine the distribution of missed associations. 
Because each candidate pair contains two strain segments, the pair-level SNR should be regarded as a summary statistic rather than a complete description of the detectability of both images.

Figure~\ref{fig:error_by_snr_large} shows the distribution of true positives and false negatives as a function of optimal matched-filter SNR. 
The SIS model has only a small number of missed associations, whereas the PM model shows a broader false-negative distribution. 
Several PM false negatives appear in intermediate-SNR bins, rather than only at the lowest SNR. 
This does not indicate that intermediate-SNR signals are intrinsically harder to classify. 
Instead, these bins can contain pairs in which the summarized SNR is moderate or high, while one image remains relatively weak or the local merger morphology is less consistent between the two branches.

Figure~\ref{fig:efficiency_sweep_combined} provides a complementary view through the SNR-binned true-positive rate. 
The SIS curve remains close to unity over most of the sampled range, while the PM curve shows stronger bin-to-bin variation. 
These variations suggest that the remaining PM errors are not governed by SNR alone, but are also affected by pairwise-consistency ambiguity, image-level SNR imbalance, and local morphology differences near merger.

Figure~\ref{fig:false_positive_local} further shows a representative false-positive pair. 
Although the two signals are not generated from the same lensed source, they show locally similar peak structures within a narrow time-sample region. 
This example illustrates how unrelated pairs can produce short local waveform similarities that lead to spurious pairwise associations.





\subsection{Ablation Study}
\label{subsec:ablation}

\begin{table}[htbp]
\centering
\caption{Ablation study of \texttt{PI-ResNet} components under ET design noise.}
\label{tab:ablation}
\renewcommand{\arraystretch}{1.15}
\setlength{\tabcolsep}{3.5pt}
\begin{tabular}{lcccc}
\tableline
\tableline
Model & \multicolumn{2}{c}{SIS} & \multicolumn{2}{c}{PM} \\
\cline{2-5}
 & Acc. & AUC & Acc. & AUC \\
\tableline
\textbf{Full} & \textbf{94.70\%} & \textbf{0.9926} & \textbf{93.90\%} & \textbf{0.9918} \\
No SE & 93.90\% & 0.9904 & 93.40\% & 0.9901 \\
No Fusion & 92.30\% & 0.9859 & 91.80\% & 0.9845 \\
\tableline
\end{tabular}
\end{table}

\begin{table}[htbp]
\centering
\caption{Cross-regime generalization performance between SIS and PM lens models under noisy ET conditions.}
\label{tab:generalization}
\renewcommand{\arraystretch}{1.2}
\begin{tabular}{llcc}
\tableline
\tableline
Training Dataset & Test Dataset & Accuracy & AUC \\
\tableline
SIS & SIS & 95.60\% & 0.9910 \\
PM  & SIS & 94.10\% & 0.9884 \\
SIS & PM  & 92.40\% & 0.9812 \\
PM  & PM  & 93.80\% & 0.9897 \\
\tableline
\end{tabular}
\end{table}

To evaluate the contribution of individual components in the PI-ResNet framework, we conduct a controlled ablation study under the ET design-noise setting. Each ablation variant was trained from scratch under the same protocol as the full model. The random seed, batch size, and training budget were kept fixed across all runs. The ablation results should therefore be interpreted as controlled relative comparisons among architectural variants, rather than as exact reproductions of the main benchmark values reported in Table~\ref{tab:comparison_semd}. The experiments isolate the effects of the Physics-Fusion mechanism and the SE blocks on both SIS and PM datasets, as summarized in Table~\ref{tab:ablation}.

Table~\ref{tab:ablation} shows that the full \texttt{PI-ResNet} architecture achieves the best overall performance on both datasets. Removing the Physics-Fusion layer leads to the largest degradation, with accuracy decreasing by about $2.40$ percentage points for SIS and $2.10$ percentage points for PM. This indicates that the explicit pairwise fusion design helps the model capture amplitude-consistency and phase-coherence relations between lensed candidate pairs under noisy conditions. Removing the SE blocks produces a smaller but consistent performance drop, suggesting that channel-wise feature recalibration provides a moderate stability improvement.

\subsection{Cross-Regime Generalization Performance}
\label{subsec:generalization}

\begin{figure}[htbp]
\centering
\includegraphics[width=\columnwidth]{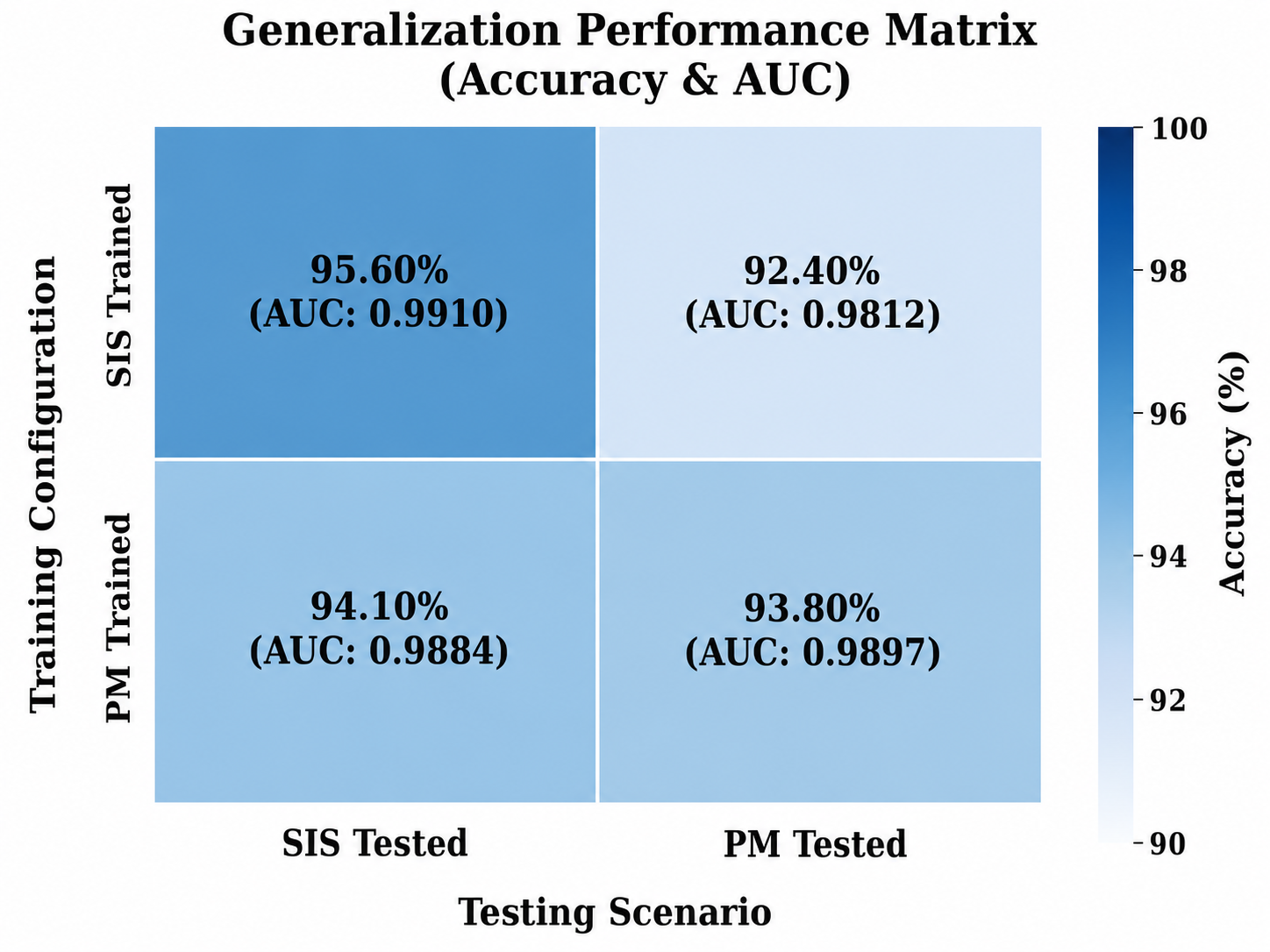}
\caption{Cross-regime generalization matrix for SIS and PM lens models under noisy ET conditions. Each entry reports the performance obtained when training on one lens model and testing on either the same or the other lens model.}
\label{fig:generalization_heatmap}
\end{figure}

To evaluate whether \texttt{PI-ResNet} captures lensing-consistency features that transfer across different lens models, we perform cross-regime tests between the SIS and PM datasets. Specifically, models trained on one lens model are evaluated both on the same lens model and on the other lens model under noisy ET conditions. The results are summarized in Table~\ref{tab:generalization}.

Table~\ref{tab:generalization} and Figure~\ref{fig:generalization_heatmap} show that \texttt{PI-ResNet} maintains relatively stable performance across SIS and PM test regimes. The same-regime evaluations provide the strongest performance, with accuracies of 95.60\% for SIS and 93.80\% for PM. In the cross-regime setting, the PM-trained model achieves 94.10\% accuracy when tested on SIS, while the SIS-trained model achieves 92.40\% accuracy when tested on PM. These results suggest that the network learns transferable pairwise lensing signatures, such as waveform-morphology consistency and phase-coherence relations, rather than relying only on lens-model-specific features.

\section{Discussion}
\label{sec:Discussion}

In this work, we show that \texttt{PI-ResNet} can extract lensing-related consistency features directly from 1D whitened detector strains, without manual feature engineering or time--frequency image construction. The results suggest that the learned representations encode pairwise consistency patterns associated with waveform morphology, magnification differences, and arrival-time or phase-evolution offsets between candidate images.

\subsection{Physical Implications of Cross-Regime Tests}
\label{subsec:physical_implications}

The cross-regime experiments show that \texttt{PI-ResNet} retains high performance when transferred between the SIS and PM datasets. As shown in Table~\ref{tab:generalization}, the PM-trained model achieves $94.10\%$ accuracy on the SIS test set, while the SIS-trained model achieves $92.40\%$ accuracy on the PM test set. These results indicate partial transferability between the two lens models, suggesting that the classifier learns pairwise lensing-consistency signatures that are not entirely specific to a single lens profile.

This interpretation should be made within the limits of the simulated parameter range. In particular, the SIS simulations in this work are restricted to a small-impact-parameter subset of the two-image regime, $0.01 \le y \le 0.3$, where lensing signatures are relatively pronounced. Therefore, the reported SIS performance should not be interpreted as a complete characterization of the full SIS parameter space. Extending the test to larger impact parameters and additional lens populations will be necessary to assess the broader generality of the learned features.

\subsection{Error Sources and SNR Dependence}

The SNR-dependent diagnostics show that missed lensed pairs are more frequent at low optimal SNR. This is expected because detector noise can obscure subtle pairwise consistency signatures, including magnification-related residual structures and phase-coherence differences. Conversely, the classifier reaches higher association efficiency once the optimal SNR becomes sufficiently large, indicating that these features are more reliably retained in stronger signals \citep{sachdev2019gstlal}.

False positives may arise when stochastic fluctuations in the simulated detector noise produce waveform-like structures after whitening and preprocessing. These fluctuations can partially mimic short-time coherent features or chirp-like morphology, increasing the probability of spurious pairwise matches. Thus, the remaining errors should be interpreted partly as a consequence of limited signal information in noisy data, rather than as purely algorithmic failures.

\subsection{Limitations and Future Outlook}

The main limitation observed in this study is sensitivity to detector-domain shift. Although \texttt{PI-ResNet} performs best under ET design noise, its performance decreases under simulated LIGO H1--L1 Gaussian-noise conditions, reaching $84.03\%$ accuracy for SIS and $78.25\%$ for PM. This indicates partial cross-detector generalization, but also shows that a model trained in one detector-noise regime may still require detector-domain adaptation before robust deployment in another PSD setting.

Future work should therefore focus on detector-domain adaptation, multi-detector consistency modeling, and validation on more realistic non-Gaussian strain data. In practical searches, \texttt{PI-ResNet} is best viewed as a fast preselection filter that can reduce the number of candidate pairs requiring computationally expensive Bayesian follow-up analyses, rather than as a replacement for full Bayesian lensing inference \citep{magare2024slick}. Extending the framework beyond the SIS and PM lens models will further test whether the 1D strain-based representation remains effective for more diverse lens populations \citep{Li_2026,cuoco2020enhancing}.

\section{Conclusion}
\label{sec:Conclusion}

In this study, we presented \texttt{PI-ResNet}, a 1D strain-based neural
framework for rapid pairwise verification of candidate lensed GW events.
The model operates directly on whitened detector strains and combines a
shared residual backbone, SE-based feature recalibration, and a
Physics-Fusion layer to model pairwise lensing-consistency signatures.

Under ET design noise, \texttt{PI-ResNet} reaches accuracies of $95.60\%$
for SIS lenses and $93.80\%$ for PM lenses. Under simulated LIGO H1--L1
Gaussian noise, the corresponding accuracies are $84.03\%$ for SIS and
$78.25\%$ for PM. These results suggest that the model retains useful
cross-detector generalization, but its performance remains sensitive to
detector-PSD domain shift. The ablation and cross-regime tests further
show that the Physics-Fusion design contributes to stable pairwise
verification across the simulated lens settings.

The proposed model provides millisecond-level pairwise screening on a standard GPU and is therefore best viewed as a fast preselection filter before more expensive Bayesian lensing analyses \citep{haris2018identifying}, rather than as a replacement for full parameter inference. Future work will focus on detector-domain adaptation, multi-detector extensions, validation on realistic non-Gaussian strain data, and more complex lens populations, including quadruply lensed systems \citep{Liao2017}.

\section*{Acknowledgments}
FZ would like to thank the MIT LIGO Laboratory for its continuous support and advice. FZ is supported by the National Natural Science Foundation of China (Grant No. 62372409) and the Ministry of Science and Technology of the People’s Republic of China (Grant No. 2023ZD0120704 of Project No. 2023ZD0120700). The authors thank Jinxuan Wu for assistance with experimental testing in this work.

\section*{Data and Code Availability}

The algorithm code and data generation scripts are publicly available at:
\url{https://github.com/zqk-7k/PI-ResNet-GW-Lensing}. For additional data access inquiries, please contact the corresponding author.

\bibliography{sample701}
\bibliographystyle{aasjournalv7}



\end{document}